\documentclass[letter,longauth]{aa} 
\usepackage{amsmath,amsfonts,amssymb}
\usepackage{url}\usepackage[varg]{txfonts}
\usepackage{graphicx}
\usepackage{enumitem} 
\usepackage{array}
\usepackage{dblfloatfix}
\usepackage{multicol}
\usepackage[normalem]{ulem} 
\usepackage{float}  % load float package first!
\usepackage{caption} 
\usepackage{pifont}
\usepackage[flushleft]{threeparttable}

\usepackage{natbib,twoopt}
\usepackage[breaklinks=true]{hyperref} %% to avoid \citeads line fills
\bibpunct{(}{)}{;}{a}{}{,}             %% natbib format for A&A and ApJ
\makeatletter
  \newcommandtwoopt{\citeads}[3][][]{\href{http://adsabs.harvard.edu/abs/#3}%
    {\def\hyper@linkstart##1##2{}%
     \let\hyper@linkend\@empty\citealp[#1][#2]{#3}}}
  \newcommandtwoopt{\citepads}[3][][]{\href{http://adsabs.harvard.edu/abs/#3}%
    {\def\hyper@linkstart##1##2{}%
     \let\hyper@linkend\@empty\citep[#1][#2]{#3}}}
  \newcommandtwoopt{\citetads}[3][][]{\href{http://adsabs.harvard.edu/abs/#3}%
    {\def\hyper@linkstart##1##2{}%
     \let\hyper@linkend\@empty\citet[#1][#2]{#3}}}
  \newcommandtwoopt{\citeyearads}[3][][]%
    {\href{http://adsabs.harvard.edu/abs/#3}
    {\def\hyper@linkstart##1##2{}%
     \let\hyper@linkend\@empty\citeyear[#1][#2]{#3}}}
\makeatother

\newcommand{\cmark}{\ding{51}}%
\newcommand{\xmark}{\ding{55}}%
\usepackage{color}
\hypersetup{colorlinks=true,linkcolor=blue,citecolor=blue,urlcolor=blue}

\usepackage[breaklinks=true]{hyperref} 

\begin{document} 

\title{HIP\,41378 observed by CHEOPS: Where is planet d?
\thanks{The CHEOPS program ID is CH\_PR110048.}
\thanks{The raw and detrended photometric time-series data are available in electronic form at the CDS via anonymous ftp to cdsarc.u-strasbg.fr (130.79.128.5) or via \url{http://cdsweb.u-strasbg.fr/cgi-bin/qcat?J/A+A/}.}
}

 \author{
S. Sulis\inst{1} $^{\href{https://orcid.org/0000-0001-8783-526X}{\includegraphics[scale=0.5]{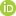}}}$, 
L. Borsato\inst{2} $^{\href{https://orcid.org/0000-0003-0066-9268}{\includegraphics[scale=0.5]{orcid.jpg}}}$, 
S. Grouffal\inst{3} $^{\href{https://orcid.org/0000-0002-2805-5869}{\includegraphics[scale=0.5]{orcid.jpg}}}$, 
H. P. Osborn\inst{4,5} $^{\href{https://orcid.org/0000-0002-4047-4724}{\includegraphics[scale=0.5]{orcid.jpg}}}$, 
A. Santerne\inst{3} $^{\href{https://orcid.org/0000-0002-3586-1316}{\includegraphics[scale=0.5]{orcid.jpg}}}$, 
A. Brandeker\inst{6} $^{\href{https://orcid.org/0000-0002-7201-7536}{\includegraphics[scale=0.5]{orcid.jpg}}}$, 
M. N. Günther\inst{7} $^{\href{https://orcid.org/0000-0002-3164-9086}{\includegraphics[scale=0.5]{orcid.jpg}}}$, 
A.~ Heitzmann\inst{8} $^{\href{https://orcid.org/0000-0002-8091-7526}{\includegraphics[scale=0.5]{orcid.jpg}}}$, 
M. Lendl\inst{8} $^{\href{https://orcid.org/0000-0001-9699-1459}{\includegraphics[scale=0.5]{orcid.jpg}}}$, 
M. Fridlund\inst{9,10} $^{\href{https://orcid.org/0000-0002-0855-8426}{\includegraphics[scale=0.5]{orcid.jpg}}}$, 
D. Gandolfi\inst{11} $^{\href{https://orcid.org/0000-0001-8627-9628}{\includegraphics[scale=0.5]{orcid.jpg}}}$, 
Y. Alibert\inst{4,12} $^{\href{https://orcid.org/0000-0002-4644-8818}{\includegraphics[scale=0.5]{orcid.jpg}}}$, 
R. Alonso\inst{13,14} $^{\href{https://orcid.org/0000-0001-8462-8126}{\includegraphics[scale=0.5]{orcid.jpg}}}$, 
T. Bárczy\inst{15} $^{\href{https://orcid.org/0000-0002-7822-4413}{\includegraphics[scale=0.5]{orcid.jpg}}}$, 
D. Barrado Navascues\inst{16} $^{\href{https://orcid.org/0000-0002-5971-9242}{\includegraphics[scale=0.5]{orcid.jpg}}}$, 
S. C. C. Barros\inst{17,18} $^{\href{https://orcid.org/0000-0003-2434-3625}{\includegraphics[scale=0.5]{orcid.jpg}}}$, 
W. Baumjohann\inst{19} $^{\href{https://orcid.org/0000-0001-6271-0110}{\includegraphics[scale=0.5]{orcid.jpg}}}$, 
T. Beck\inst{12}, 
W. Benz\inst{12,4} $^{\href{https://orcid.org/0000-0001-7896-6479}{\includegraphics[scale=0.5]{orcid.jpg}}}$, 
M. Bergomi\inst{20} $^{\href{https://orcid.org/ 0000-0001-7564-2233}{\includegraphics[scale=0.5]{orcid.jpg}}}$, 
N. Billot\inst{8} $^{\href{https://orcid.org/0000-0003-3429-3836}{\includegraphics[scale=0.5]{orcid.jpg}}}$, 
A. Bonfanti\inst{19} $^{\href{https://orcid.org/0000-0002-1916-5935}{\includegraphics[scale=0.5]{orcid.jpg}}}$, 
C. Broeg\inst{12,4} $^{\href{https://orcid.org/0000-0001-5132-2614}{\includegraphics[scale=0.5]{orcid.jpg}}}$, 
A. Collier Cameron\inst{21} $^{\href{https://orcid.org/0000-0002-8863-7828}{\includegraphics[scale=0.5]{orcid.jpg}}}$, 
C. Corral van Damme\inst{7}, 
A. C. M. Correia\inst{22}, 
Sz. Csizmadia\inst{23} $^{\href{https://orcid.org/0000-0001-6803-9698}{\includegraphics[scale=0.5]{orcid.jpg}}}$, 
P. E. Cubillos\inst{24,19}, 
M. B. Davies\inst{25} $^{\href{https://orcid.org/0000-0001-6080-1190}{\includegraphics[scale=0.5]{orcid.jpg}}}$, 
M. Deleuil\inst{1} $^{\href{https://orcid.org/0000-0001-6036-0225}{\includegraphics[scale=0.5]{orcid.jpg}}}$, 
A. Deline\inst{8}, 
L. Delrez\inst{26,27,28} $^{\href{https://orcid.org/0000-0001-6108-4808}{\includegraphics[scale=0.5]{orcid.jpg}}}$, 
O. D. S. Demangeon\inst{17,18} $^{\href{https://orcid.org/0000-0001-7918-0355}{\includegraphics[scale=0.5]{orcid.jpg}}}$, 
B.-O. Demory\inst{4,12} $^{\href{https://orcid.org/0000-0002-9355-5165}{\includegraphics[scale=0.5]{orcid.jpg}}}$, 
A. Derekas\inst{29}, 
B. Edwards\inst{30}, 
D. Ehrenreich\inst{8,31} $^{\href{https://orcid.org/0000-0001-9704-5405}{\includegraphics[scale=0.5]{orcid.jpg}}}$, 
A. Erikson\inst{23}, 
A. Fortier\inst{12,4} $^{\href{https://orcid.org/0000-0001-8450-3374}{\includegraphics[scale=0.5]{orcid.jpg}}}$, 
L. Fossati\inst{19} $^{\href{https://orcid.org/0000-0003-4426-9530}{\includegraphics[scale=0.5]{orcid.jpg}}}$, 
K. Gazeas\inst{32}, 
M. Gillon\inst{26} $^{\href{https://orcid.org/0000-0003-1462-7739}{\includegraphics[scale=0.5]{orcid.jpg}}}$, 
M. Güdel\inst{33}, 
Ch. Helling\inst{19,34}, 
S. Hoyer\inst{1} $^{\href{https://orcid.org/0000-0003-3477-2466}{\includegraphics[scale=0.5]{orcid.jpg}}}$, 
K. G. Isaak\inst{7} $^{\href{https://orcid.org/0000-0001-8585-1717}{\includegraphics[scale=0.5]{orcid.jpg}}}$, 
L. L. Kiss\inst{35,36}, 
J. Korth\inst{37} $^{\href{https://orcid.org/0000-0002-0076-6239}{\includegraphics[scale=0.5]{orcid.jpg}}}$,  
K. W. F. Lam\inst{23} $^{\href{https://orcid.org/0000-0002-9910-6088}{\includegraphics[scale=0.5]{orcid.jpg}}}$, 
J. Laskar\inst{38} $^{\href{https://orcid.org/0000-0003-2634-789X}{\includegraphics[scale=0.5]{orcid.jpg}}}$, 
A. Lecavelier des Etangs\inst{39} $^{\href{https://orcid.org/0000-0002-5637-5253}{\includegraphics[scale=0.5]{orcid.jpg}}}$, 
D. Magrin\inst{2} $^{\href{https://orcid.org/0000-0003-0312-313X}{\includegraphics[scale=0.5]{orcid.jpg}}}$, 
P. F. L. Maxted\inst{40} $^{\href{https://orcid.org/0000-0003-3794-1317}{\includegraphics[scale=0.5]{orcid.jpg}}}$, 
C. Mordasini\inst{12,4}, 
V. Nascimbeni\inst{2} $^{\href{https://orcid.org/0000-0001-9770-1214}{\includegraphics[scale=0.5]{orcid.jpg}}}$, 
G. Olofsson\inst{6} $^{\href{https://orcid.org/0000-0003-3747-7120}{\includegraphics[scale=0.5]{orcid.jpg}}}$, 
R. Ottensamer\inst{33}, 
I. Pagano\inst{41} $^{\href{https://orcid.org/0000-0001-9573-4928}{\includegraphics[scale=0.5]{orcid.jpg}}}$, 
E. Pallé\inst{13,14} $^{\href{https://orcid.org/0000-0003-0987-1593}{\includegraphics[scale=0.5]{orcid.jpg}}}$, 
G. Peter\inst{42} $^{\href{https://orcid.org/0000-0001-6101-2513}{\includegraphics[scale=0.5]{orcid.jpg}}}$, 
D. Piazza\inst{43}, 
G. Piotto\inst{2,44} $^{\href{https://orcid.org/0000-0002-9937-6387}{\includegraphics[scale=0.5]{orcid.jpg}}}$, 
D. Pollacco\inst{45}, 
D. Queloz\inst{46,47} $^{\href{https://orcid.org/0000-0002-3012-0316}{\includegraphics[scale=0.5]{orcid.jpg}}}$, 
R. Ragazzoni\inst{2,44} $^{\href{https://orcid.org/0000-0002-7697-5555}{\includegraphics[scale=0.5]{orcid.jpg}}}$, 
N. Rando\inst{7}, 
H. Rauer\inst{23,48} $^{\href{https://orcid.org/0000-0002-6510-1828}{\includegraphics[scale=0.5]{orcid.jpg}}}$, 
I. Ribas\inst{49,50} $^{\href{https://orcid.org/0000-0002-6689-0312}{\includegraphics[scale=0.5]{orcid.jpg}}}$, 
N. C. Santos\inst{17,18} $^{\href{https://orcid.org/0000-0003-4422-2919}{\includegraphics[scale=0.5]{orcid.jpg}}}$, 
G. Scandariato\inst{41} $^{\href{https://orcid.org/0000-0003-2029-0626}{\includegraphics[scale=0.5]{orcid.jpg}}}$, 
D. Ségransan\inst{8} $^{\href{https://orcid.org/0000-0003-2355-8034}{\includegraphics[scale=0.5]{orcid.jpg}}}$, 
A. E. Simon\inst{12,4} $^{\href{https://orcid.org/0000-0001-9773-2600}{\includegraphics[scale=0.5]{orcid.jpg}}}$, 
A. M. S. Smith\inst{23} $^{\href{https://orcid.org/0000-0002-2386-4341}{\includegraphics[scale=0.5]{orcid.jpg}}}$, 
S. G. Sousa\inst{17} $^{\href{https://orcid.org/0000-0001-9047-2965}{\includegraphics[scale=0.5]{orcid.jpg}}}$, 
M. Stalport\inst{27,26}, 
M. Steinberger\inst{19}, 
Gy. M. Szabó\inst{51,52} $^{\href{https://orcid.org/0000-0002-0606-7930}{\includegraphics[scale=0.5]{orcid.jpg}}}$, 
A. Tuson\inst{53} $^{\href{https://orcid.org/0000-0002-2830-9064}{\includegraphics[scale=0.5]{orcid.jpg}}}$, 
S. Udry\inst{8} $^{\href{https://orcid.org/0000-0001-7576-6236}{\includegraphics[scale=0.5]{orcid.jpg}}}$, 
S. Ulmer-Moll\inst{8} $^{\href{https://orcid.org/0000-0003-2417-7006}{\includegraphics[scale=0.5]{orcid.jpg}}}$, 
V. Van Grootel\inst{27} $^{\href{https://orcid.org/0000-0003-2144-4316}{\includegraphics[scale=0.5]{orcid.jpg}}}$, 
J. Venturini\inst{8} $^{\href{https://orcid.org/0000-0001-9527-2903}{\includegraphics[scale=0.5]{orcid.jpg}}}$, 
E. Villaver\inst{13,14}, 
N. A. Walton\inst{53} $^{\href{https://orcid.org/0000-0003-3983-8778}{\includegraphics[scale=0.5]{orcid.jpg}}}$, 
T. G. Wilson\inst{45} $^{\href{https://orcid.org/0000-0001-8749-1962}{\includegraphics[scale=0.5]{orcid.jpg}}}$, 
D. Wolter\inst{23}, 
T. Zingales\inst{44,2} $^{\href{https://orcid.org/0000-0001-6880-5356}{\includegraphics[scale=0.5]{orcid.jpg}}}$
}
 
 \institute{    
\label{inst:1} Aix Marseille Univ, CNRS, CNES, LAM, 38 rue Frédéric Joliot-Curie, 13388 Marseille, France \and
\label{inst:2} INAF, Osservatorio Astronomico di Padova, Vicolo dell'Osservatorio 5, 35122 Padova, Italy \and
\label{inst:3} Universit\'e Aix Marseille, CNRS, CNES, LAM, Marseille, France \and
\label{inst:4} Center for Space and Habitability, University of Bern, Gesellschaftsstrasse 6, 3012 Bern, Switzerland \and
\label{inst:5} Department of Physics and Kavli Institute for Astrophysics and Space Research, Massachusetts Institute of Technology, Cambridge, MA 02139, USA \and
\label{inst:6} Department of Astronomy, Stockholm University, AlbaNova University Center, 10691 Stockholm, Sweden \and
\label{inst:7} European Space Agency (ESA), European Space Research and Technology Centre (ESTEC), Keplerlaan 1, 2201 AZ Noordwijk, The Netherlands \and
\label{inst:8} Observatoire astronomique de l'Université de Genève, Chemin Pegasi 51, 1290 Versoix, Switzerland \and
\label{inst:9} Leiden Observatory, University of Leiden, PO Box 9513, 2300 RA Leiden, The Netherlands \and
\label{inst:10} Department of Space, Earth and Environment, Chalmers University of Technology, Onsala Space Observatory, 439 92 Onsala, Sweden \and
\label{inst:11} Dipartimento di Fisica, Università degli Studi di Torino, via Pietro Giuria 1, I-10125, Torino, Italy \and
\label{inst:12} Weltraumforschung und Planetologie, Physikalisches Institut, University of Bern, Gesellschaftsstrasse 6, 3012 Bern, Switzerland \and
\label{inst:13} Instituto de Astrofísica de Canarias, Vía Láctea s/n, 38200 La Laguna, Tenerife, Spain \and
\label{inst:14} Departamento de Astrofísica, Universidad de La Laguna, Astrofísico Francisco Sanchez s/n, 38206 La Laguna, Tenerife, Spain \and
\label{inst:15} Admatis, 5. Kandó Kálmán Street, 3534 Miskolc, Hungary \and
\label{inst:16} Depto. de Astrofísica, Centro de Astrobiología (CSIC-INTA), ESAC campus, 28692 Villanueva de la Cañada (Madrid), Spain \and
\label{inst:17} Instituto de Astrofisica e Ciencias do Espaco, Universidade do Porto, CAUP, Rua das Estrelas, 4150-762 Porto, Portugal \and
\label{inst:18} Departamento de Fisica e Astronomia, Faculdade de Ciencias, Universidade do Porto, Rua do Campo Alegre, 4169-007 Porto, Portugal \and
\label{inst:19} Space Research Institute, Austrian Academy of Sciences, Schmiedlstrasse 6, A-8042 Graz, Austria \and
\label{inst:20} INAF - Osservatorio Astronomico di Padova \and
\label{inst:21} Centre for Exoplanet Science, SUPA School of Physics and Astronomy, University of St Andrews, North Haugh, St Andrews KY16 9SS, UK \and
\label{inst:22} CFisUC, Department of Physics, University of Coimbra, 3004-516 Coimbra, Portugal \and
\label{inst:23} Institute of Planetary Research, German Aerospace Center (DLR), Rutherfordstrasse 2, 12489 Berlin, Germany \and
\label{inst:24} INAF, Osservatorio Astrofisico di Torino, Via Osservatorio, 20, I-10025 Pino Torinese To, Italy \and
\label{inst:25} Centre for Mathematical Sciences, Lund University, Box 118, 221 00 Lund, Sweden \and
\label{inst:26} Astrobiology Research Unit, Université de Liège, Allée du 6 Août 19C, B-4000 Liège, Belgium \and
\label{inst:27} Space sciences, Technologies and Astrophysics Research (STAR) Institute, Université de Liège, Allée du 6 Août 19C, 4000 Liège, Belgium \and
\label{inst:28} Institute of Astronomy, KU Leuven, Celestijnenlaan 200D, 3001 Leuven, Belgium \and
\label{inst:29} ELTE Gothard Astrophysical Observatory, 9700 Szombathely, Szent Imre herceg u. 112, Hungary \and
\label{inst:30} SRON Netherlands Institute for Space Research, Niels Bohrweg 4, 2333 CA Leiden, Netherlands \and
\label{inst:31} Centre Vie dans l’Univers, Faculté des sciences, Université de Genève, Quai Ernest-Ansermet 30, 1211 Genève 4, Switzerland \and
\label{inst:32} National and Kapodistrian University of Athens, Department of Physics, University Campus, Zografos GR-157 84, Athens, Greece \and
\label{inst:33} Department of Astrophysics, University of Vienna, Türkenschanzstrasse 17, 1180 Vienna, Austria \and
\label{inst:34} Institute for Theoretical Physics and Computational Physics, Graz University of Technology, Petersgasse 16, 8010 Graz, Austria \and
\label{inst:35} Konkoly Observatory, Research Centre for Astronomy and Earth Sciences, 1121 Budapest, Konkoly Thege Miklós út 15-17, Hungary \and
\label{inst:36} ELTE E\"otv\"os Lor\'and University, Institute of Physics, P\'azm\'any P\'eter s\'et\'any 1/A, 1117 Budapest, Hungary \and
\label{inst:37} Lund Observatory, Division of Astrophysics, Department of Physics, Lund University, Box 118, 22100 Lund, Sweden \and
\label{inst:38} IMCCE, UMR8028 CNRS, Observatoire de Paris, PSL Univ., Sorbonne Univ., 77 av. Denfert-Rochereau, 75014 Paris, France \and
\label{inst:39} Institut d'astrophysique de Paris, UMR7095 CNRS, Université Pierre \& Marie Curie, 98bis blvd. Arago, 75014 Paris, France \and
\label{inst:40} Astrophysics Group, Lennard Jones Building, Keele University, Staffordshire, ST5 5BG, United Kingdom \and
\label{inst:41} INAF, Osservatorio Astrofisico di Catania, Via S. Sofia 78, 95123 Catania, Italy \and
\label{inst:42} Institute of Optical Sensor Systems, German Aerospace Center (DLR), Rutherfordstrasse 2, 12489 Berlin, Germany \and
\label{inst:43} Physikalisches Institut, University of Bern, Sidlerstrasse 5, 3012 Bern, Switzerland \and
\label{inst:44} Dipartimento di Fisica e Astronomia "Galileo Galilei", Università degli Studi di Padova, Vicolo dell'Osservatorio 3, 35122 Padova, Italy \and
\label{inst:45} Department of Physics, University of Warwick, Gibbet Hill Road, Coventry CV4 7AL, United Kingdom \and
\label{inst:46} ETH Zurich, Department of Physics, Wolfgang-Pauli-Strasse 2, CH-8093 Zurich, Switzerland \and
\label{inst:47} Cavendish Laboratory, JJ Thomson Avenue, Cambridge CB3 0HE, UK \and
\label{inst:48} Institut fuer Geologische Wissenschaften, Freie Universitaet Berlin, Maltheserstrasse 74-100,12249 Berlin, Germany \and
\label{inst:49} Institut de Ciencies de l'Espai (ICE, CSIC), Campus UAB, Can Magrans s/n, 08193 Bellaterra, Spain \and
\label{inst:50} Institut d’Estudis Espacials de Catalunya (IEEC), Gran Capità 2-4, 08034 Barcelona, Spain \and
\label{inst:51} ELTE E\"otv\"os Lor\'and University, Gothard Astrophysical Observatory, 9700 Szombathely, Szent Imre h. u. 112, Hungary \and
\label{inst:52} HUN-REN--ELTE Exoplanet Research Group, Szent Imre h. u. 112., Szombathely, H-9700, Hungary \and
\label{inst:53} Institute of Astronomy, University of Cambridge, Madingley Road, Cambridge, CB3 0HA, United Kingdom
}

\authorrunning{S. Sulis et al.}

 \date{accepted to A\&A}

%%%%%%%%%%%%%%%%%%%%%%%%%%%%%%%%%%%%%%%%
% ABSTRACT AND KEYWORDS
%%%%%%%%%%%%%%%%%%%%%%%%%%%%%%%%%%%%%%%%

% 5 {} token are mandatory
 
  \abstract
   {
   HIP 41378 d is a long-period planet that has only been observed to transit twice, three years apart, with K2. According to stability considerations and a partial detection of the Rossiter-McLaughlin effect,  $P_\mathrm{d} = 278.36$~d has been determined to be the most likely orbital period. We targeted HIP 41378 d with CHEOPS at the predicted transit timing based on $P_\mathrm{d}= 278.36$~d, but the observations show no transit.
  {We find that large ($>22.4$ hours) transit timing variations (TTVs) could explain this non-detection during the CHEOPS observation window.}
   We also investigated the possibility of an incorrect orbital solution, which would have major implications for our knowledge of this system. If $P_\mathrm{d} \neq 278.36$~d, the periods that minimize the eccentricity would be $101.22$~d and $371.14$~d.  
   The shortest orbital period will be tested by TESS, which will observe HIP 41378 in Sector 88 starting in January 2025.
    Our study shows the importance of a mission like CHEOPS, which today is the only mission able to make long observations (i.e., from space) to track the ephemeris of long-period planets possibly affected by large TTVs.
    }


   \maketitle
%
%-------------------------------------------------------------------

\section{Introduction}

\begin{table*}[t!]
    \centering
    \caption{Log file of CHEOPS observations: file keys referring to the files name in the CHEOPS database, starting date (BJD), total observation duration, number of data points before and after detrending, number of CHEOPS orbits, duty cycle (DC), and exposure time.}
    \begin{tabular}{ccccccc}
\hline
File key                    & Starting date & Duration & Data points & \# orbits & DC & Exposure time \\
in the CHEOPS database   & (BJD)         & (h)      &   (raw/detrended)     &     &   (\%)     &  (s) \\
\hline
CH\_PR110048\_TG032601\_V0300  & $2459949.183$ & $32.3$  &  $1838$ / $1827$ & $20$ & $61$ & $38$ \\
\hline
    \end{tabular}
    \label{tab_cheops}
\end{table*}
 
The bright F-type star HIP\,41378 ($m_V \approx 8.9$; $T_\mathrm{eff}=6290 \pm 77$ K; \citeads{2019AJ....158..248L}) is transited by at least five exoplanets \citepads{2016ApJ...827L..10V} with very long orbital periods, up to 1.5 years for planet f \citep[][]{2019arXiv191107355S}. With this period, the transit probability is as low as 0.4~\%. {The planetary system transiting HIP\,41378 is thus a unique laboratory for studying the compositions and atmospheres of planets that are not extremely irradiated, and which fall in the gap between the cold outer Solar System planets and hot gazeous planets}. It is also an interesting system to search for moons and rings{, in particular around planet f (see, e.g.,} \citeads{2020A&A...635L...8A}; \citeads{2023A&A...675A.174S}; {\citeads{2022AJ....163..277B}; \citeads{2022ApJ...927L...5A}; \citeads{2023ApJS..269...31E}; \citeads{2023AJ....166..208H}). }

In this system, planet d has been seen to transit twice, three years apart, in photometric data from the K2 mission, during Campaign~5 \citepads{2016ApJ...827L..10V} and Campaign 18 (\citeads{2019AJ....157..185B}; \citeads{2019AJ....157...19B}). This resulted in $23$ possible period aliases: namely, $3$ years and all the harmonics down to about $50$ days, which is the size of the continuous window of observations. Combining the long transit duration ($t_{\mathrm{dur}} \approx 12.5$ hours) and the precise and accurate density of the host star derived thanks to asteroseismology, \citetads{2019AJ....158..248L} estimate that the most likely orbital period for planet d that minimizes the eccentricity in this massive multi-planet system is $P_\mathrm{d}\,=\,278.36\pm 0.001$~d.

\begin{figure}[t!]
\centering
\resizebox{\hsize}{!}{\includegraphics{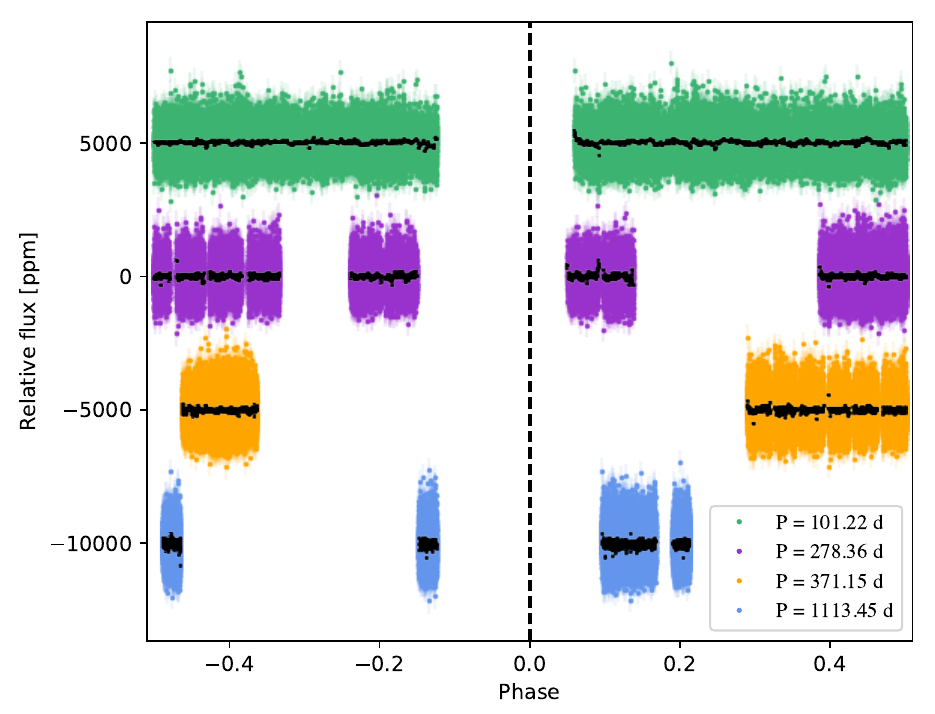}}
\caption{TESS observations of HIP\,41378 (sectors 7, 34, 44, 45, 46, 61, and 72) phase-folded at different orbital periods (see the legend). $3$-hour-binned data are shown in black. The different series have been shifted in flux for visibility. The transit duration is within the thickness of the vertical line.}
\label{fig_tess}  
\end{figure}

The star HIP\,41378 was also intensively observed via radial velocity (RV), mainly with the HARPS \citepads{2003Msngr.114...20M} and HARPS-N \citepads{2012SPIE.8446E..1VC} spectrographs. {However, no significant RV signal was found at $\sim278$~d by  \citet{2019arXiv191107355S}; possibly because the signal may have been too faint for the stability of these instruments (small periodicities at $\sim350-410$ days are, however, reported in their RV data analyses).} The star was recently photometrically monitored by the TESS space telescope over seven sectors, but no transit of planet d was detected (see Fig.~\ref{fig_tess}) despite an expected S/N above 400. These TESS observations allowed $16$ of the $23$ orbital period aliases inferred after the K2 campaign for planet d to be discarded (see \citeads{2022A&A...668A.172G}); the value $P_\mathrm{d}=278.36$ d remained the most plausible.

Given its long duration and shallow depth (650 ppm), the transit is challenging to detect using ground-based facilities. Instead, \citetads{2022A&A...668A.172G} used high-resolution spectroscopy and detected a signal compatible with an egress of the Rossiter-McLaughlin (RM) effect (\citeads{1893AstAp..12..646H};  \citeads{1924ApJ....60...15R}; \citeads{1924ApJ....60...22M}) at the expected transit time and with an amplitude compatible with the $\sim$278 d period, providing evidence in support of this orbital solution.
 
Assuming the orbital period of $\sim$278~d is accurate, we observed HIP 41378 with CHEOPS \citep{2021ExA....51..109B} at the expected transit timing of planet d ($T_{0,\mathrm{d}} = 2 459 949.8787 \pm 0.008$; \citeads{2022A&A...668A.172G}), which represented the only transit event visible in both the nominal and first extended CHEOPS mission.
With this, we aimed to confirm the planet's orbital period and to constrain possible transit-timing variations (TTVs) of HIP 41378 d, as observed for planet f \citepads{2021MNRAS.504L..45B}.  
These objectives are in line with results from other CHEOPS programs that have confirmed moderately long-period planets, for example TOI-2076c \citepads{2022A&A...664A.156O}, HIP 9618c \citepads{2023MNRAS.523.3069O}, TOI-5678b \citepads{2023A&A...674A..43U}, HD22946d \citepads{2023A&A...674A..44G}, HD15906b and c \citepads{2023MNRAS.523.3090T}, and TOI-815c \citepads{2024A&A...685A...5P}.
This Letter reports on the CHEOPS observations of HIP 41378 and discusses their implications for the system characterization.

%--------------------------------------------------------------------
\section{CHEOPS observations and analysis}
\label{sec_data}

\begin{figure}[t!]
\centering
\resizebox{\hsize}{!}{\includegraphics{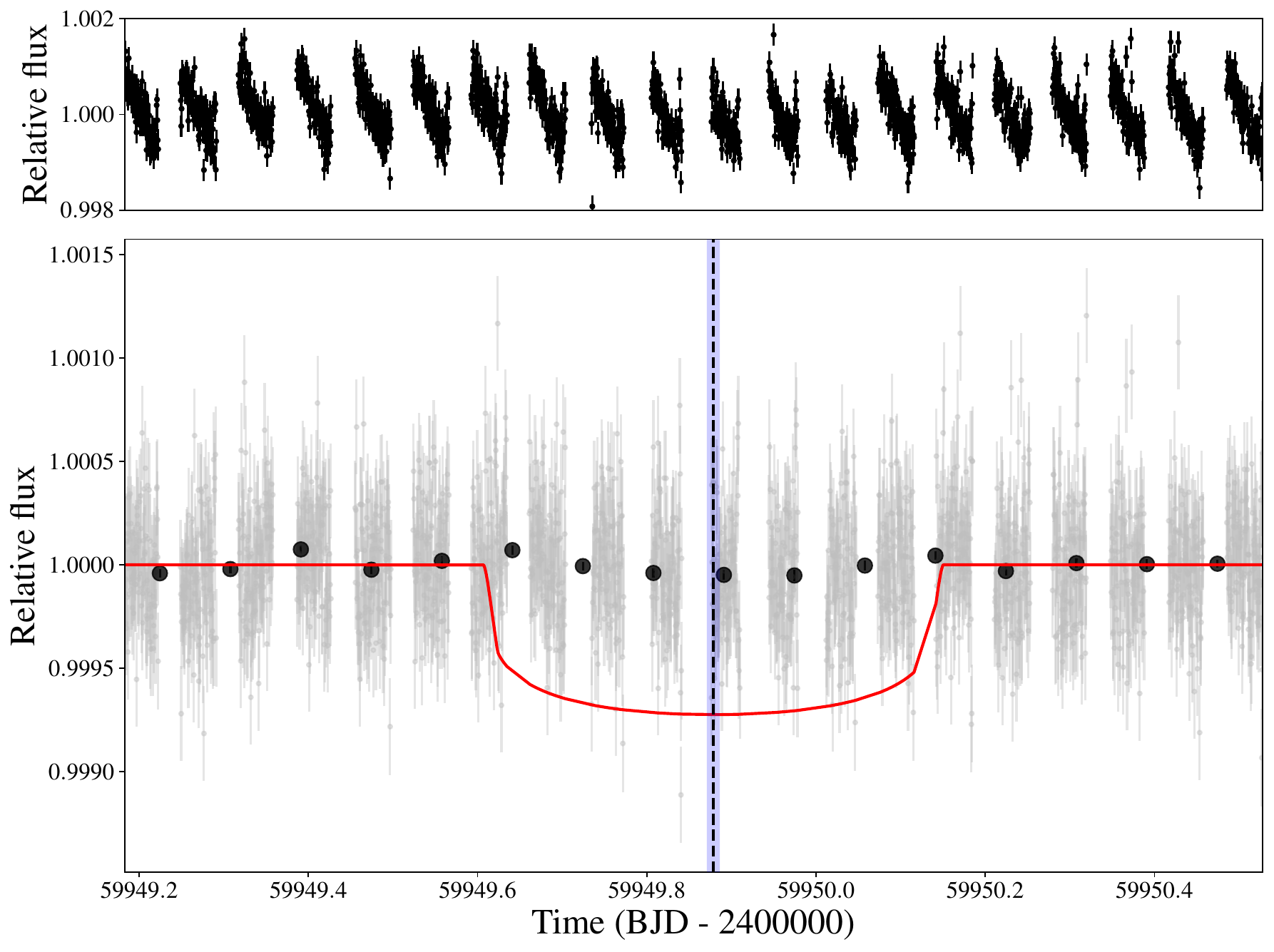}}
\caption{CHEOPS observations of HIP\,41378 at the predicted time of planet d's transit. \textit{Top:} {Raw light curve.} \textit{Bottom:} Detrended (gray) and $2$-hour-binned light curves (black). The transit model is in red. The dotted vertical line indicates the predicted time of mid-transit $T_{0,\mathrm{d}}$, assuming an orbital period of $P_\mathrm{d}=278.36$ d. The shaded blue region represents the uncertainties on $T_{0,\mathrm{d}}$. }
\label{fig_cheops}  
\end{figure}

The CHEOPS visit of HIP 41378 was scheduled as part of the guaranteed time observing program. HIP\,41378 was observed for $\mathrm{T_{WIN}} = 32.3$ hours from January 4, 2023, at T16:14:16 to January 6, 2023, at T00:32:21 UTC. The integration time of each measuring point was $\sim$38~s.
During these observations, the duty cycle was close to $61\%$. The origin of the data gaps lies in the low-Earth orbit of CHEOPS: when the satellite crosses the South Atlantic Anomaly, when the Earth occults the target, or when stray light contamination is too high, data are not downloaded \citepads{2021ExA....51..109B}.  {A summary of CHEOPS HIP\,41378 observations is presented in Table~\ref{tab_cheops}.}
Observations were processed with the automatic CHEOPS Data Reduction Pipeline (DRP; v14.1.2), described in \citetads{2020A&A...635A..24H}. We adopted the target flux obtained with the default radius aperture, which is $r=25$ pixels. Two main contaminants are present in this photometric aperture, but they are faint (G-mag\,=\,14 and 18.2) compared to the target (G-mag\,=\,8.81).
The ``raw'' light curve, obtained by removing outliers exceeding $5\sigma$ ($11$ data points), is presented in the top panel of Fig.~\ref{fig_cheops}.

We detrended the light curve from instrumental and environmental noise using the python package \texttt{pycheops}\footnote{\url{ https://github.com/pmaxted/pycheops} {(version 1.1.7)}} \citepads{2022MNRAS.514...77M}. 
{
To de-correlate the light curve from CHEOPS systematics, we first identified the relevant parameters using the \textit{should\_I\_decorr()} function of \texttt{pycheops}. 
This function fits various combinations of trends between flux, time, roll angle ($\Phi$), and the different variables provided by DRP (e.g., target centroid positions, background, contamination flux, and smear contamination). {For each combination, it calculates the Bayesian information criterion (BIC) and determines the combination that yields the lowest BIC value} (see Section 2 of \citealt{2022MNRAS.514...77M} for details). 
{We find that the best-fitting combination of detrending parameters involves} the first- and second-order derivative in time (\texttt{dfdt} and \texttt{d2fdt2}),
the first- and second-order derivative of the offset in the x and y centroid positions (\texttt{dfdx}, \texttt{d2fdx2}, \texttt{dfdy}, and \texttt{d2fdy2}),
the first three harmonics of the roll angle (in $\cos{\Phi}$ and $\sin{\Phi}$), 
and linear trends with the sky background (\texttt{dfdbg}), contamination (\texttt{dfdcontam}), and smearing systematics (\texttt{dfdsmear}).
}
The root mean square of the light curve before and after this de-correlation step is $523$ and $271$ ppm, respectively. %{; and the raw and detrended CHEOPS data are available through CDS.}
The final light curve is shown in Fig.~\ref{fig_cheops}. No transit is detected at the predicted ephemeris $T_{0,\mathrm{d}}$ calculated by \citetads{2022A&A...668A.172G} assuming an orbital period of $P_\mathrm{d}=278.36$ d {and a transit depth of $650$ ppm \citepads{2016ApJ...827L..10V}.}  

%--------------------------------------------------------------------------------- 
\section{Discussion: Where is planet d?}
\label{sec_dis}

We devised two main hypotheses to explain the non-detection of the transit of HIP\,41378\,d in the CHEOPS light curve. They are as follows.

\subsection{Possible large TTV signal}\label{ssec_ttv}

\begin{figure}[t]
\centering
\resizebox{\hsize}{!}{\includegraphics{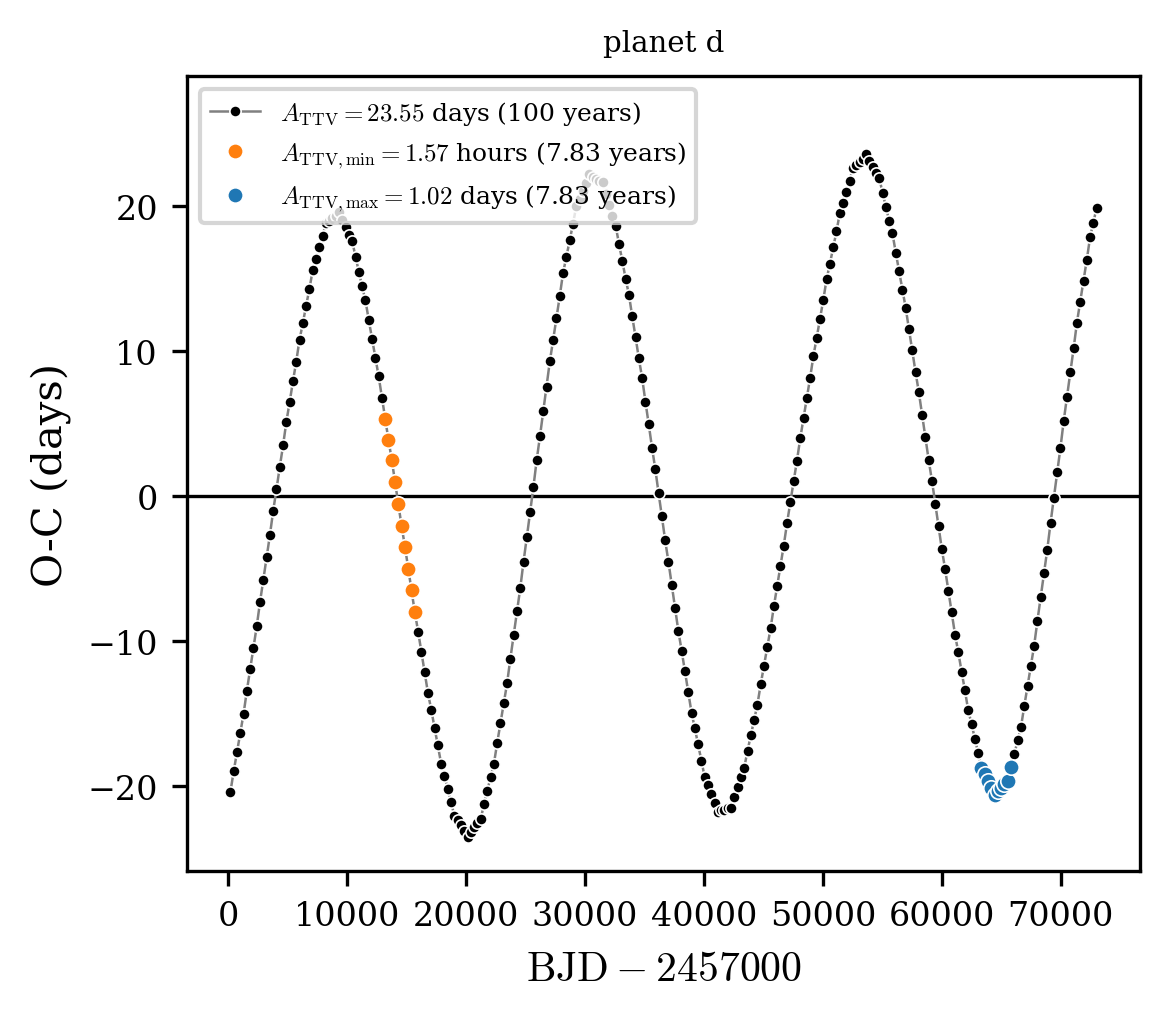}}
\caption{
$O-C$ diagram of synthetic transit times ($T_{0,\mathrm{s}}$) of 100~years (black circles) of dynamical simulations with \texttt{TRADES} computed using the \citet{2019arXiv191107355S} solution and including all planets.
Highlighted are the two observing windows of $\sim7.8$~years whose re-computed $O-C$ (as described in Section~\ref{ssec_ttv}) show the minimum (orange) TTV amplitude ($A_\mathrm{TTV,min}=1.57$~hours)
and the maximum (blue) TTV amplitude ($A_\mathrm{TTV,max}=1.02$~days).
}
\label{fig_OCd}
\end{figure}

Assuming the alias at $P_\mathrm{d}\sim278$~d and the \citetads{2019arXiv191107355S} solution are correct, we dynamically integrated the whole system with  \textsc{TRADES}\footnote{\url{https://github.com/lucaborsato/trades}.} \citepads{Borsato2014A&A...571A..38B, Borsato2019MNRAS.484.3233B, Borsato2021MNRAS.506.3810B}.
The solution in \citetads{2019arXiv191107355S} was the result of the sum of noninteractive Keplerian orbits, so it provides a mean orbital configuration without knowledge of the exact position of all the planets at some reference time.

Under these conditions, we had to take the values of the initial parameters based on the
\citet{2019arXiv191107355S} solution.
We calculated $\mathcal{M} = \frac{2\pi}{P}(t_\mathrm{ref}-\tau),$ where 
$t_\mathrm{ref}$ is the transit time of planet d in Campaign 18 of K2 and 
$\tau$ is the time of passage at pericenter.
For planet d we assumed $\mathcal{M}_\mathrm{d}(t_\mathrm{ref}) = 0\degr$,
and for the other five planets we assigned random values between $0\degr$ and $360\degr$.
We assumed the argument of the pericenters ($\omega$) to be $90\degr$ for all the planets.
We integrated the orbits of the six planets for 200 years and extracted the synthetic transit times ($T_{0,\mathrm{s}}$) from the simulation.
We exploited the possible TTV signal by constructing an observed minus calculated ($O-C$) diagram: the $O$ is the synthetic $T_{0,\mathrm{s}}$ (the initial reference time is the median $T_{0}$ and the period is the median of the difference of successive $T_{0,\mathrm{s}}$), and the $C$ is a linear ephemeris computed on the synthetic $T_{0,\mathrm{s}}$.
The amplitude of the synthetic TTV, $A_\mathrm{TTV}$, computed as the semi-amplitude of the $O-C$, is approximately $23.55$~days over a temporal baseline of 200~years (see Figure~\ref{fig_OCd}).
However, our observational baseline covers only $7.8$~years (from K2 to CHEOPS).\ Therefore, we started with a temporal window spanning $7.8$~years and shifted it by one transit at a time; by selecting the $T_{0, \mathrm{s}}$ that fall in that time range, we recomputed the $O-C$ after refitting the linear ephemeris.
The $A_\mathrm{TTV}$ was derived for each moving window, revealing that the potential TTV signal in the observation windows of the \citet{2019arXiv191107355S} orbital configuration can vary from a minimum of $A_\mathrm{TTV,min} \sim 2$~hours to a maximum of $A_\mathrm{TTV,max} \sim 1$~day (see an example in Fig.~\ref{fig_OCd}).

\par

To reduce computational time, we decided to simulate only the three outer planets (d, e, and f), 
repeating the orbital integration and sliding TTV. This approach yielded comparable results.
So we assumed that the effect of the three inner planets is negligible for the purpose of our sliding TTV analysis.
We ran $10\, 000$ simulations with \textsc{TRADES}, varying the parameters based on the uncertainty
and upper bounds\footnote{
We could not sample from the posterior (which would have allowed us to conserve the covariance between parameters) because
it was no longer available.\ We had to rely on the marginalized parameter values and uncertainties
reported in \citet{2019arXiv191107355S}.} proposed by \citet{2019arXiv191107355S}.
We set the lower limit of masses to $0.1\, M_{\oplus}$ and forced only positive eccentricities.
The argument of pericenters ($\omega$) of all planets and 
the mean anomaly ($\mathcal{M}$) of planets e and f
were assigned random values between $0\degr$ and $360\degr$,
the $\mathcal{M}_\mathrm{d}$ was fixed to $0\degr$,
and the longitude of the ascending node ($\Omega$) was set to $180\degr$ for all planets.
For each simulation, we integrated for 200 years and computed the sliding TTV.
Out of $10\, 000$ simulations, $7\, 921$ ($79.21\%$) were stable over 200 years.
Of these, $3\, 438$ ($43.40\%$ of stable simulations and $34.38\%$ of all the simulations) had at least one TTV window above the required threshold of $(\mathrm{T_{WIN}} + t_{\mathrm{dur}})/2 = 22.4$~hours.

This shows that $P_\mathrm{d}=278.36$~d could be the true orbital period of planet d, despite the lack of a transit in the CHEOPS light curve. 
This is compatible with the RM detection reported in \citetads{2022A&A...668A.172G}. 
However, it would imply that our CHEOPS observations were carried out during a particular configuration of the system with {an $A_\mathrm{TTV}>22.4$~hours}.

\subsection{A different period for planet d}

\begin{figure}[t!]
\centering
\resizebox{\hsize}{!}{\includegraphics{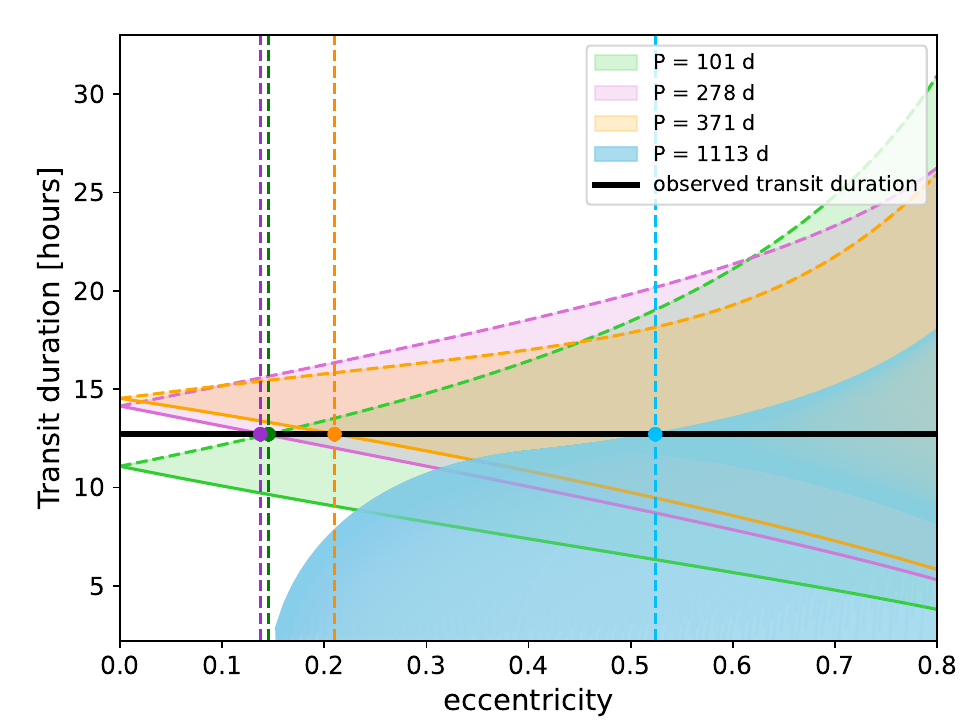}}
\caption{Transit duration for eccentricities from 0 to 0.8 for HIP41378 d for four orbital periods (see the legend). The dotted vertical lines represent the minimal eccentricity for each orbital period to obtain the observed transit duration of $12.5$ hours (thick black horizontal line). }
\label{fig_eccentricities}  
\end{figure}

The second scenario is an incorrect orbital period solution for HIP\,41378\,d.
Combining the results of the seven TESS sectors and the new result obtained by CHEOPS, three solutions are possible for the planetary orbital period, as presented in Table~\ref{tab:OrbitalSolution}: $101.22$ d, $371.15$ d, and $1113.44$ d. None of these three solutions are compatible with the RM detection reported in \citetads{2022A&A...668A.172G}. %, which is not understood. 
 Many sources of error might have led to a false detection of the RM effect (e.g., instrumental systematics, stellar variability, and distortion of the RV after the detrending of the data). If the orbital period of $278.36$~d turns out to be wrong, it would be interesting to understand the origin of the RM-like signal observed by \citetads{2022A&A...668A.172G} for future RM analyses of this iconic system.
%}

For each of these periods, we estimated the minimal eccentricity required to observe a transit duration of $t_\mathrm{dur}$\,$\approx$\,12.5~hours using Eqs. (7) and (14) from \citetads{2010exop.book...55W}. 
We used the values from \citetads{2019arXiv191107355S} for the radius and inclination. We used both the formula for the transit duration from \citetads{2010exop.book...55W} and that from \citetads{2010MNRAS.407..301K}, and we obtained similar results. 
The expected transit duration for the four possible orbital periods as a function of the eccentricities and for $\omega$ ranging from periastron to apastron is represented in Fig. \ref{fig_eccentricities}.
The period that minimizes the eccentricity is $P_\mathrm{d} = 278.36$ d, which was one of the reasons this orbital period was considered the most plausible. This period is compatible with \citet{2019AJ....158..248L}'s analysis. An orbital period of $101.22$ d with the planet seen at apastron also gives a low eccentricity similar to the one for $P_\mathrm{d} = 278.36 $ d. Period $P_\mathrm{d} = 371.15$ d is also compatible with a reasonable minimal eccentricity of $\sim$0.21. Only period $1113.44$ d seems excluded as a high eccentricity is necessary ($\sim0.52$). We note the different behavior of the transit duration for $P_\mathrm{d} =1113.44$ d in Fig. \ref{fig_eccentricities}, which is explained by impact parameters becoming close to or larger than $1$ for some eccentricity values in Eq. (7) of \citetads{2010exop.book...55W}.

HIP 41378 is planned to be observed again by TESS during Sector 88 (from January 14 to February 11, 2025), for which a transit of planet d might be detected only if $P_\mathrm{d} = 101.22$ d. 
If $P_\mathrm{d} = 278.36$, the next opportunity to detect a transit with CHEOPS will be in January 2026. If $P_\mathrm{d} = 371.14$ d or $1113.44$~d, there is no opportunity to observe any transit in the next $15$ years with HST, TESS, CHEOPS, or ARIEL.

These new possible period estimates change our view of the HIP\,41378 planetary system. Planet~e was observed only once in transit during K2's C5 campaign, with a transit duration of $\sim$13~hours. Using a global analysis of K2 transits with RV data, \citet{2019arXiv191107355S} found an orbital period of $P_\mathrm{e} = 369 \pm 10$~d. Using HIP\,41378d's new orbital period constraints, the period of planet e may be different (especially if $P_\mathrm{d} = 371.14$~d or $1113.44$~d).
Detecting additional transits of planets e and d is necessary to understand these long-period planets. 

\begin{table}[t!]
    \centering

\caption{Summary of the remaining orbital periods for HIP41378~d (assuming no TTV). }
    
    \begin{tabular}{|c|c|c|c|c|}
    \hline
        Orbital period [d] & TESS & CHEOPS &  RM & $e_\mathrm{min}$\\
    \hline
    $1113.4465 \pm 0.0034$ & \cmark & \cmark & \xmark & $\sim 0.52$  \\ 
    $371.1488 \pm 0.0011$ & \cmark & \cmark & \xmark  & $\sim 0.21$\\
    $278.3616 \pm 0.0009$ & \cmark  & \xmark & \cmark & $\sim 0.14$\\
    $101.2224 \pm 0.0003$ & \cmark & \cmark &  \xmark & $\sim 0.15$  \\
    \hline
    \end{tabular}

    \begin{tablenotes}
        \small
        \item Orbital solutions compatible with no transit in TESS and CHEOPS observations are shown with a check mark (\cmark), while those incompatible are shown with an \xmark. Orbital period values are taken from \citetads{2019AJ....157...19B}. The RM column corresponds to the results in \citetads{2022A&A...668A.172G} for the RM detection of the transit. The last column displays the minimal eccentricity that corresponds to $t_\mathrm{dur}$\,$\approx$\,12.5~hours. 
    \end{tablenotes}
    
    \label{tab:OrbitalSolution}
\end{table}

%-------------------------------------- 
\section{Conclusions}
\label{ccl}

We observed HIP\,41378 with CHEOPS for $32.3$ hours to refine the ephemeris of planet d and constrain the TTVs in order to dynamically analyze the whole system \citepads{2019arXiv191107355S}. Planet d was first seen to transit twice with K2 
% (\citeads{2016ApJ...827L..10V}); \citeads{2019AJ....157..185B}; \citeads{2019AJ....157...19B})
\citep{2016ApJ...827L..10V,2019AJ....157..185B,2019AJ....157...19B}. 
TESS observations enable one to rule out several period aliases computed after the K2 campaign and are compatible with four period aliases, namely $P_\mathrm{d} \sim 101, ~278, ~371$, and $1113$ d. Of these values, $P_\mathrm{d} \sim278$ d was believed to be the most likely period when\ taking the RM observations by \citetads{2022A&A...668A.172G} into account.
 
CHEOPS observations, planned at $P_\mathrm{d} \sim 278$~d, showed no transit. The expected S/N of this event rules out a missed planet detection with CHEOPS due to instrumental systematics. Two possible scenarios are discussed in this Letter.

The first scenario assumes that the planet orbital period of $P_\mathrm{d} \sim 278$~d is correct but that the transit was not seen due to a large TTV {($>22.4$ hours)}. With our simulations, we indeed find configurations where {$A_\mathrm{TTV}>22.4$~hours}.
Detecting the planet from ground-based observations (with windows of $<10$ hours) would therefore be challenging. 
We note that the current TTV analysis could benefit from significant refinement from an updated orbital solution of this multi-planet system. We also emphasize the importance of considering dynamical constraints in future RV analyses of this system, which are needed to assign a robust probability to the TTV scenario.

The second scenario assumes that the period solution of $\sim 278$ d is incorrect. This would not be consistent with the results of \citetads{2022A&A...668A.172G} and would mean that  \citetads{2022A&A...668A.172G} misinterpreted the RM-like signal. If we consider a different period for planet d, this would change the derived period of the mono-transit of planet e. A complete reanalysis of the RV data is needed, but this goes beyond the scope of this Letter. 

In conclusion, CHEOPS observations do not allow us to conclude which of the four possible orbital periods for planet d is correct. However, two periods (namely $P_\mathrm{d} \sim 101$~d and $371$~d) seem to be possible alternatives as they minimize the planet's eccentricity and would imply that the CHEOPS observations were made in {the $A_\mathrm{TTV}<22.4$~hours configuration}. Future observations by TESS (Sector 88) will enable us to draw conclusions about the $P_\mathrm{d}\sim 101$~d period.

We note that other factors could explain this non-detection, such as the variation in the inclination, which could make the transit event disappear on a short timescale (e.g., from \citeads{2023A&A...675A.174S}, the precession period would be $\sim 1000-2500$ years). Such a scenario would be interesting to investigate when a more reliable orbital solution for the system becomes available.

This study shows the importance of monitoring long-period planets to keep track of and refine their  ephemerides, which may be affected by large TTVs. It may then be necessary to carry out long-duration observations (i.e., from space) around the predicted planet ephemeris to secure the transit detection. 
The CHEOPS mission, designed to characterize known exoplanets, is particularly well suited to this task. With the future PLATO space mission (scheduled for 2026; \citeads{2014ExA....38..249R}), the discovery of long-period planets should intensify. Consequently, extending the CHEOPS mission post-PLATO launch and/or considering the development of additional CHEOPS-like missions, potentially stationed at L2 (to reduce pointing constraints and time gaps), would be fully justified.

%\newpage

\begin{acknowledgements}
The authors thank the anonymous referee for her/his helpful comments that improved this letter.
% CHEOPS
CHEOPS is an ESA mission in partnership with Switzerland with important contributions to the payload and the ground segment from Austria, Belgium, France, Germany, Hungary, Italy, Portugal, Spain, Sweden, and the United Kingdom. The CHEOPS Consortium would like to gratefully acknowledge the support received by all the agencies, offices, universities, and industries involved. Their flexibility and willingness to explore new approaches were essential to the success of this mission. CHEOPS data analysed in this article will be made available in the CHEOPS mission archive (\url{https://cheops.unige.ch/archive_browser/}). 
% Sophia
SS and MD acknowledge support from CNES, as well as the Programme National de Planétologie (PNP) of CNRS-INSU. 
% From python file
LBo, GBr, VNa, IPa, GPi, RRa, GSc, VSi, and TZi acknowledge support from CHEOPS ASI-INAF agreement n. 2019-29-HH.0. 
This work has been carried out within the framework of the NCCR PlanetS supported by the Swiss National Science Foundation under grants 51NF40\_182901 and 51NF40\_205606. 
ABr was supported by the SNSA. 
MNG is the ESA CHEOPS Project Scientist and Mission Representative, and as such also responsible for the Guest Observers (GO) Programme. MNG does not relay proprietary information between the GO and Guaranteed Time Observation (GTO) Programmes, and does not decide on the definition and target selection of the GTO Programme. 
ML acknowledges support of the Swiss National Science Foundation under grant number PCEFP2\_194576. 
MF and CMP gratefully acknowledge the support of the Swedish National Space Agency (DNR 65/19, 174/18). 
DG gratefully acknowledges financial support from the CRT foundation under Grant No. 2018.2323 “Gaseousor rocky? Unveiling the nature of small worlds”. 
YAl acknowledges support from the Swiss National Science Foundation (SNSF) under grant 200020\_192038. 
We acknowledge financial support from the Agencia Estatal de Investigación of the Ministerio de Ciencia e Innovación MCIN/AEI/10.13039/501100011033 and the ERDF “A way of making Europe” through projects PID2019-107061GB-C61, PID2019-107061GB-C66, PID2021-125627OB-C31, and PID2021-125627OB-C32, from the Centre of Excellence “Severo Ochoa” award to the Instituto de Astrofísica de Canarias (CEX2019-000920-S), from the Centre of Excellence “María de Maeztu” award to the Institut de Ciències de l’Espai (CEX2020-001058-M), and from the Generalitat de Catalunya/CERCA programme. 
We acknowledge financial support from the Agencia Estatal de Investigación of the Ministerio de Ciencia e Innovación MCIN/AEI/10.13039/501100011033 and the ERDF “A way of making Europe” through projects PID2019-107061GB-C61, PID2019-107061GB-C66, PID2021-125627OB-C31, and PID2021-125627OB-C32, from the Centre of Excellence “Severo Ochoa'' award to the Instituto de Astrofísica de Canarias (CEX2019-000920-S), from the Centre of Excellence “María de Maeztu” award to the Institut de Ciències de l’Espai (CEX2020-001058-M), and from the Generalitat de Catalunya/CERCA programme. 
S.C.C.B. acknowledges support from FCT through FCT contracts nr. IF/01312/2014/CP1215/CT0004. 
C.B. acknowledges support from the Swiss Space Office through the ESA PRODEX program. 
ACC acknowledges support from STFC consolidated grant numbers ST/R000824/1 and ST/V000861/1, and UKSA grant number ST/R003203/1. 
ACMC ackowledges support from the FCT, Portugal, through the CFisUC projects UIDB/04564/2020 and UIDP/04564/2020, with DOI identifiers 10.54499/UIDB/04564/2020 and 10.54499/UIDP/04564/2020, respectively. 
P.E.C. is funded by the Austrian Science Fund (FWF) Erwin Schroedinger Fellowship, program J4595-N. 
This project was supported by the CNES. 
The Belgian participation to CHEOPS has been supported by the Belgian Federal Science Policy Office (BELSPO) in the framework of the PRODEX Program, and by the University of Liège through an ARC grant for Concerted Research Actions financed by the Wallonia-Brussels Federation. 
L.D. thanks the Belgian Federal Science Policy Office (BELSPO) for the provision of financial support in the framework of the PRODEX Programme of the European Space Agency (ESA) under contract number 4000142531. 
This work was supported by FCT - Funda\c{c}\~{a}o para a Ci\^{e}ncia e a Tecnologia through national funds and by FEDER through COMPETE2020 through the research grants UIDB/04434/2020, UIDP/04434/2020, 2022.06962.PTDC. 
O.D.S.D. is supported in the form of work contract (DL 57/2016/CP1364/CT0004) funded by national funds through FCT. 
B.-O. D. acknowledges support from the Swiss State Secretariat for Education, Research and Innovation (SERI) under contract number MB22.00046. 
This project has received funding from the Swiss National Science Foundation for project 200021\_200726. It has also been carried out within the framework of the National Centre of Competence in Research PlanetS supported by the Swiss National Science Foundation under grant 51NF40\_205606. The authors acknowledge the financial support of the SNSF. 
M.G. is an F.R.S.-FNRS Senior Research Associate. 
CHe acknowledges support from the European Union H2020-MSCA-ITN-2019 under Grant Agreement no. 860470 (CHAMELEON). 
S.H. gratefully acknowledges CNES funding through the grant 837319.
J.K. gratefully acknowledges the support of the Swedish Research Council (VR: Etableringsbidrag 2017-04945). 
{A.C., A.D., B.E., K.G., and J.K. acknowledge their role as ESA-appointed CHEOPS Science Team Members.} 
K.W.F.L. was supported by Deutsche Forschungsgemeinschaft grants RA714/14-1 within the DFG Schwerpunkt SPP 1992, Exploring the Diversity of Extrasolar Planets. 
This work was granted access to the HPC resources of MesoPSL financed by the Region Ile de France and the project Equip@Meso (reference ANR-10-EQPX-29-01) of the programme Investissements d'Avenir supervised by the Agence Nationale pour la Recherche. 
PM acknowledges support from STFC research grant number ST/R000638/1. 
This work was also partially supported by a grant from the Simons Foundation (PI Queloz, grant number 327127). 
NCSa acknowledges funding by the European Union (ERC, FIERCE, 101052347). Views and opinions expressed are however those of the author(s) only and do not necessarily reflect those of the European Union or the European Research Council. Neither the European Union nor the granting authority can be held responsible for them. 
A. S. acknowledges support from the Swiss Space Office through the ESA PRODEX program. 
S.G.S. acknowledge support from FCT through FCT contract nr. CEECIND/00826/2018 and POPH/FSE (EC). 
The Portuguese team thanks the Portuguese Space Agency for the provision of financial support in the framework of the PRODEX Programme of the European Space Agency (ESA) under contract number 4000142255. 
GyMSz acknowledges the support of the Hungarian National Research, Development and Innovation Office (NKFIH) grant K-125015, a a PRODEX Experiment Agreement No. 4000137122, the Lendulet LP2018-7/2021 grant of the Hungarian Academy of Science and the support of the city of Szombathely. 
V.V.G. is an F.R.S-FNRS Research Associate. 
JV acknowledges support from the Swiss National Science Foundation (SNSF) under grant PZ00P2\_208945. 
NAW acknowledges UKSA grant ST/R004838/1. 
TWi acknowledges support from the UKSA and the University of Warwick. 
\end{acknowledgements}

%%%%%%%%%%%%%%%%%%%%%%%%%%%%%%%%%%%%%%%%
% BIBLIOGRAPHY
%%%%%%%%%%%%%%%%%%%%%%%%%%%%%%%%%%%%%%%%

\bibliographystyle{aa} 
\bibliography{bibtex}

\end{document}